\documentclass[prl,twocolumn,showpacs,preprintnumbers,amsmath,amssymb]{revtex4}

\usepackage{graphicx}% Include figure files
\usepackage{bm}% bold math
\usepackage{amsmath}
\usepackage{latexsym}

\newcommand{\be}{\begin{equation}}
\newcommand{\ee}{\end{equation}}
\newcommand{\req}[1]{Eq.~(\ref{#1})}

\newcommand{\rfig}[1]{Fig.~\ref{#1}}
\newcommand{\rref}[1]{Ref.~\cite{#1}}

\begin{document}

\title{The epitaxial-graphene/graphene-oxide junction, \\an essential step towards epitaxial graphene electronics}

\author{Xiaosong Wu$^1$, Mike Sprinkle$^1$, Xuebin Li$^1$, Fan Ming$^1$, Claire Berger$^{1,2}$, Walt A. de Heer$^{1}$}
\affiliation{
$^1$School of Physics, Georgia Institute of Technology, Atlanta, GA 30332 \\
$^2$CNRS-Institut N\'{e}el, BP 166, 38042 Grenoble, France}

\date{\today}

\begin{abstract}
Graphene oxide (GO) flakes have been deposited to bridge the gap between two epitaxial graphene electrodes to produce all-graphene devices. Electrical measurements indicate the presence of Schottky barriers (SB) at the graphene/graphene oxide junctions, as a consequence of the band-gap in GO. The barrier height is found to be about 0.7 eV, and is reduced after annealing at 180 $^\circ$C, implying that the gap can be tuned by changing the degree of oxidation. A lower limit of the GO mobility was found to be 850 cm$^2$/Vs, rivaling silicon. {\it In situ} local oxidation of patterned epitaxial graphene has been achieved.
\end{abstract}


\maketitle

Inspired by the exceptional properties of carbon nanotubes, epitaxial graphene based electronics was conceived as a possible new platform for post-CMOS electronics. In contrast to carbon nanotubes, graphene layers can be patterned to produce interconnected all-carbon structures, thereby overcoming a wide variety of problems facing nanotube-based electronics. Our earlier work focused primarily on producing and characterizing device quality epitaxial graphene (EG) on silicon carbide \cite{Berger2004, Berger2006,Wu2007,deHeer2007,Footnote1}. Here we demonstrate the production and properties of the epitaxial-graphene/graphene-oxide Schottky barrier. We also successfully chemically patterned epitaxial graphene to produce seamless graphene oxide to graphene junctions, thereby dramatically enhancing epitaxial graphene electronics.

We recently showed that EG can be reliably patterned over large areas to produce hundreds of functioning high mobility field effect transistors (FET) over the entire surface of a 3$\times$4 mm chip using high $k$ dielectrics \cite{Kedzierski2007}. Next steps involve patterning and tailoring the properties of EG. Conventional semiconductor devices rely on a significant band gap; graphene, by contrast, is a semimetal, which severely limits the switching potential of graphene FETs (currently the maximum off-to-on resistance ratio for EG is about 35). The high mobility of EG (up to 25,000 cm$^2$/Vs) offsets this deficiency for certain specialized applications. Clearly, the versatility of graphene electronics is greatly increased by converting graphene into a semiconductor. One way to achieve this is by nanopatterning. It was predicted that the electronic structure of a nanoscopic graphene ribbon should mimic that of a carbon nanotube \cite{Nakada1996, Wakabayashi1999} and semiconducting nanopatterned graphene ribbons on exfoliated graphene flakes have been demonstrated \cite{Han2007a,Li2008}.

A far more convenient scheme is to chemically convert graphene to a semiconductor. In this Letter we demonstrate the properties of (semiconducting) graphene oxide (GO), integrated into patterned EG structures. GO, first described in 1859 \cite{Brodie1859}, consists of graphene layers whose surfaces are oxidized without disrupting the hexagonal graphene topology. Impressive demonstrations of deposited single layer GO \cite{Kovtyukhova1999} spurred research into alternative methods to produce a single graphene layer, by reducing deposited GO back to graphene \cite{Stankovich2007, NanoL2007}. In contrast, here we are interested in the semiconducting properties of GO and the capability to locally convert EG to GO. For electronics applications, multilayered epitaxial graphene has several advantages over single layer graphene: the patterned structures are more robust, the interior layers are protected from the environment, and the layered structure allows intercalation.

Suspensions of $\sim1 \mu$m GO flakes were obtained from Mallouk et al. \cite{Kovtyukhova1999}. An ac dielectrophoresis method was used to deposit flakes over pairs of Au electrodes patterned on an oxidized Si wafer, or over patterned EG electrodes, separated by 400, 800, or 1400 nm gaps. We found that an ac voltage of 2-3 volts peak-to-peak at 20-50 kHz produced optimal results. Samples were finally heated to 100 $^\circ$C for 30 minutes in order to drive off absorbed water vapor.

\begin{figure}
\includegraphics[width=0.36\textwidth]{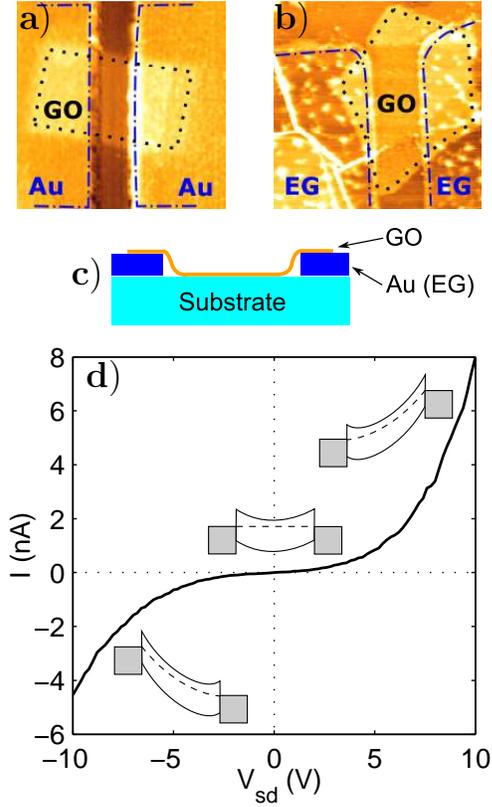}
\caption{\label{IV} EG/GO metal-semiconductor-metal (MSM) device. {\bf a}) A bilayer rectangular GO flake (outlined by a black dotted line) over a 400nm Au gap. The Au pads are outlined by blue dash-dot lines. {\bf b}) A pentagonal GO flake bridges two EG electrodes. Both images are $2 \mu$m$\times 2 \mu$m. The bright spots on EG are e-beam resist (PMMA) residue, while the bright lines are wrinkles that are often seen in C-face EG. {\bf c}) The layout of GO devices (side view). {\bf d}) $I$-$V$ characteristics of an 800 nm device, consisting of back-to-back Schottky diodes. The inset schematically shows band diagrams for the device under various biasing conditions. The asymmetry of the $I$-$V$ characteristics reflects dissimilarities of the two junctions.}
\end{figure}

We have taken AFM images of over 30 GO flakes spanning electrode gaps. Most of them are single layers that are usually flat and free of wrinkles. The measured thickness of a single GO layer on SiO$_2$ ranges from 0.9 to 1.5 nm, consistent with \rref{Kovtyukhova1999}, while it is 1.5 to 2.2 nm on EG. \rfig{IV}a and \ref{IV}b show typical images of flakes over pairs of electrodes. \rfig{IV}b shows a single layer GO flake deposited over an EG gap. All single layer flakes have remarkably similar $I$-$V$ characteristics, as discussed in detail below. \rfig{IV}a shows a bilayer flake (indicated by the step on the right electrode) deposited over a Au electrode pair separated by a 400 nm gap. Unlike single layer flakes, the bilayer flake is insulating for bias voltages up to 20 V, which may indicate that the electronic properties of bilayer GO differ significantly from those of single layer GO. Devices made on Au electrodes and on EG electrodes exhibit similar current-voltage ($I$-$V$) behavior, characteristic of back-to-back Schottky diodes (see below).

We have measured devices with varying gap widths: 400 nm, 800 nm, 1400 nm. A typical $I$-$V$ curve of an EG-GO-EG device, shown in \rfig{IV}d, exhibits strong nonlinearity. The $I$-$V$s do not systematically vary with the gap width. Because bulk resistance would scale with the applied electric field (not the potential), the nonlinearity is not an indication of bulk resistance in GO. This specifically rules out strong localization effects in GO as the origin of the nonlinearity \cite{Hill1971}.

Another important feature is the asymmetry of the $I$-$V$ with respect to the bias voltage. This often-seen asymmetry correlates with the ratio of the lengths of the contact edges on the two electrodes (not the area of the two GO/contact overlap regions). The asymmetry, and its correlation with the length of the edge, indicates that the impedance is primarily due to the contact edge between one of the two EG electrodes and the GO flake ({\it i.e.} the junction length). As with carbon nanotube SBs \cite{Heinze2002}, this picture is consistent with a SB at the GO/conductor edge, and inconsistent with an impedance distributed over the contact area of the GO and the electrode. Therefore, the structures correspond to two back-to-back SBs (\rfig{IV}d). When a bias voltage is applied, one SB is under reverse bias, while the other is forward-biased. Consequently, the impedance will always be dominated by characteristics of the reverse-biased SB. The impedance of the reverse-biased SB should be approximately inversely proportional to the junction length, {\it i.e.} the lower impedance branch of an $I$-$V$ corresponds to the reverse-biased SB with the longer junction length. With this insight, the polarity of the SBs is determined and the carrier type can be identified. We find that some GO flakes are p-type, while others are n-type and carrier densities are rather low and variable (order of $10^{10}-10^{11}$ cm$^{-2}$, see below). The arbitrary nature of the carrier type and density indicates that environment and substrate effects play a role. This situation is similar to carbon nanotubes and exfoliated graphene \cite{Kong2000,Tan2007}, which are prepared under similar, non-pristine conditions, causing arbitrary doping by impurities.

A detailed analysis of the SB characteristics follows. The SB at the interface of a 2-dimensional electron gas (2DEG) from a modulation doped heterostructure and 3-dimensional metal has recently been studied. Based on a thermionic emission model, the quasisaturation reverse bias current through a SB can be described by \cite{Anwar1999}:
\be
J_s=\frac{2q}{h^2}\sqrt{2\pi m^*}(k_BT)^\frac{3}{2} \times \exp\left( -\frac{q\phi_b}{k_BT} \right ) \times \exp\left( \frac{q\Delta\phi_b}{k_BT} \right )
\label{current}
\ee
where $m^*$ is the effective electron mass, $\phi_b$ is the effective SB height. $\Delta\phi_b$ is the field dependence of the barrier. For image-force lowering of the barrier, it follows \cite{Rhoderick1988}:
\be
\Delta\phi_b=\left\{ \frac{q^3N_d}{8\pi^2(\epsilon_s')^2\epsilon_s} \left( \phi_b+V_r-\xi-\frac{k_BT}{q} \right)  \right\}^\frac{1}{4}
\label{barrierlowering}
\ee
where $N_d$ is the ionized donor (or acceptor) density, $V_r$ is the reverse bias voltage, $\xi$ is the distance between the Fermi level of the semiconductor and the bottom of the conduction band. $\epsilon'$ and $\epsilon$ are high frequency and static dielectric constants of the semiconductor, respectively. $\phi_b$ is usually less than 1 eV and $k_BT\approx30$ meV at $T=300$ K, so at high bias, the image force lowering of the barrier is proportional to the one-fourth power of the bias.

\begin{figure}
\includegraphics[width=0.43\textwidth]{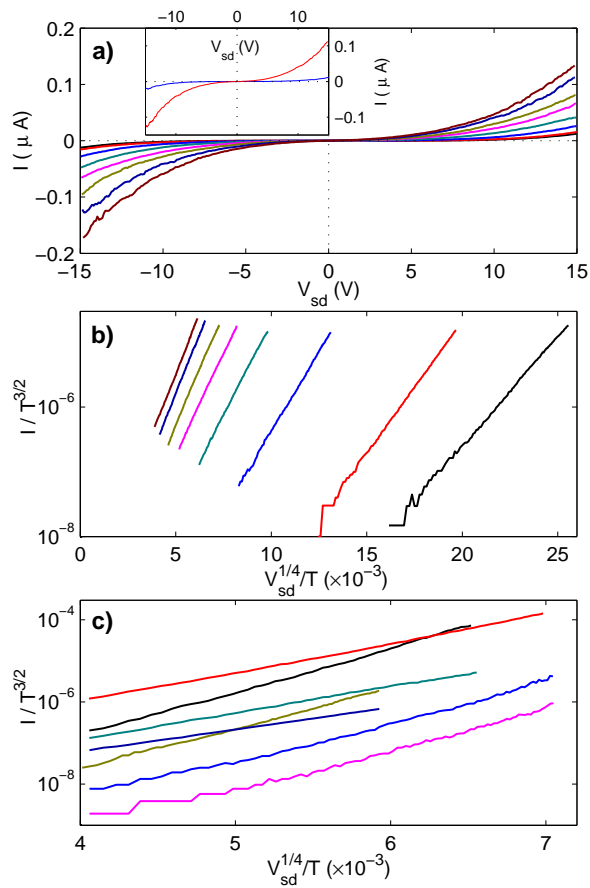}
\caption{\label{Thermionic} {\bf a}) and {\bf b}), $I$-$V$ characteristics of a 400 nm device at several temperatures: 77 (black), 100 (red), 150 (blue), 200 (cyan), 240 (pink), 270 (dark yellow), 300 (dark blue) and 320 K (magenta). The sample was cured at 180 $^\circ$C for 16 hours. {\bf a}) Nonlinear $I$-$V$. Inset: $I$-$V$ before (blue) and after (red) curing. The increased current indicates a lowering of the SB height. {\bf b}) $I/T^{3/2}$ as a function of $V_{sd}^{1/4}/T$ for $V_{sd}>2$ V. The observed linear dependence is as expected for a back-biased Schottky diode where the current is determined by thermionic emission over the barrier (\req{current}, (\ref{barrierlowering})). The slope of the line gives the ionized donor density. {\bf c}) $I/T^{3/2}$ vs $V_{sd}^{1/4}/T$ plots at 300K for two Au/GO devices (black and red), the device in {\bf a}) before annealing (blue), another 400 nm EG/GO device before and after annealing (cyan and pink), an 800 nm EG/GO device (dark yellow) and a 1400 nm EG/EG device (dark blue).}
\end{figure}

\begin{figure}
\includegraphics[width=0.44\textwidth]{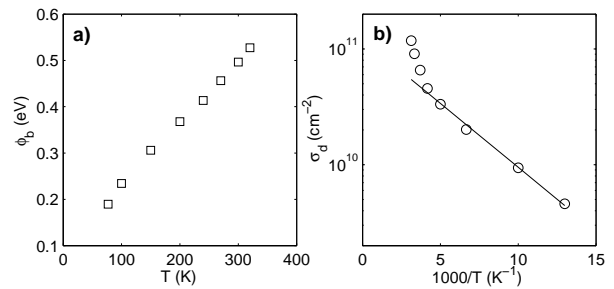}
\caption{\label{Tdependence} The temperature dependence of the device parameters. {\bf a}) The temperature dependence of the SB height. {\bf b}) The area density of ionized donors as a function of inverse temperature. Circles: experiment; Line: A fit to $N_d\propto\exp(-E_i/2k_BT)$ gives $E_i\approx61$ meV.}
\end{figure}

Figure \ref{Thermionic}a and \ref{Thermionic}b show the $I$-$V$ characteristics of a 400 nm device at 77, 100, 150, 200, 240, 270, 300 and 320 K. As the temperature decreases, the current is suppressed, as expected for a thermionic emission current. To test if \req{current} and (\ref{barrierlowering}) describe the data, we plot $I/T^{3/2}$ as a function of $V_{sd}^{1/4}/T$ in a semi-log plot. The curves are linear, which is consistent with \req{current}. The barrier height of the SB was calculated from the intercept, assuming the mass of a free electron. This mass approximation is justified by the fact that the estimation of the barrier height weakly depends on $m^*$ ({\it e. g.}, a change of two orders of magnitude in $m^*$ results in a change of less than 10\% in $\phi_b$). The barrier height at room temperature is estimated to be 0.5 eV; it decreases with temperature (\rfig{Tdependence}a). Such temperature dependence of the barrier height is commonly seen in SB diodes and is associated with the ideality factor $n$ of the diode \cite{Wagner1983} or a temperature dependent band gap. Effects that can cause a departure from unity in the value of $n$ include thermionic-field emission processes, interface effects, electron-hole recombination or nonuniformities in the SB \cite{Rhoderick1988}. Further study is needed to identify the effect in this case.

The ionized donor (or acceptor) density $N_d$ was calculated from the slope of $\ln(I/T^{3/2})$ versus $V_{sd}^{1/4}/T$. The dielectric constant can be approximated as an average of that of the substrate SiC and that of the air above, {\it i.e.} $\epsilon=(\epsilon_\text{SiC}+1)/2$. Since $\epsilon_\text{SiC}=6.7$ and $\epsilon'_\text{SiC}=10$, we have $\epsilon=3.85$ and $\epsilon'=5.5$. According to \req{barrierlowering}, we find that $N_d\approx4.5\times10^{17}$ cm$^{-3}$ at 300 K, corresponding to an area density $\sigma_d$ of $9.1\times10^{10}$ cm$^{-2}$, assuming that the thickness of the flake is 2 nm. For donors with an ionization energy $E_i$, $N_d\propto\exp(-E_i/2k_BT)$ when $E_i\gg k_BT$ \cite{Kittel1995}. Consequently, $E_i$ of the sample was obtained from the temperature dependence of $N_d$ (\rfig{Tdependence}b). We find $E_i\approx61$ meV, a typical value for these SBs. More than twenty samples have been studied, and all are described by \req{current} (see representative data in \rfig{Thermionic}c). The doping density was found to be between $2.2\times10^{10}$ and $6.1\times10^{11}$ cm$^{-2}$, while $\phi_b$ ranges from 0.45 to 0.7 eV.

It is known that GO loses oxygen when heated above 100 $^\circ$C \cite{Stankovich2007}, {\it i.e.} the degree of oxidization can be adjusted by curing. The data in \rfig{Thermionic} were obtained after a curing process at 180 $^\circ$C for 16 hours. The $I$-$V$ characteristics before and after curing are plotted in the inset of \rfig{Thermionic}a. A significant increase of the current after curing was observed and analysis reveals that the doping density decreased (from $3.8\times10^{11}$ to $9.1\times10^{10}$ cm$^{-2}$). More importantly, $\phi_b$ decreased from 0.7 to 0.5 eV. A similar trend was observed in all other cured samples. This indicates that we can use thermal oxygen desorption to tune the band structure, as suggested in \rref{Lee2005}.

\begin{figure}
\includegraphics[width=0.43\textwidth]{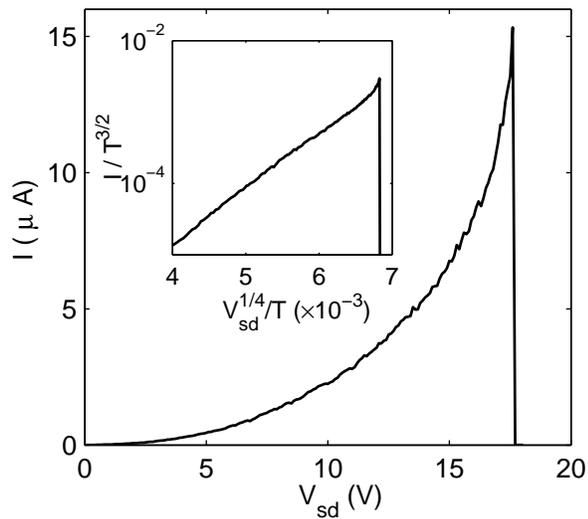}
\caption{\label{Burn} An Au-GO-Au device reached its breakdown voltage and burned out. The insets is a plot of $I/T^{3/2}$ as a function of $V_{sd}^{1/4}/T$ for $V_{sd}>2$ V.}
\end{figure}

Since the impedance of the device is dominated by the SB, we cannot directly measure the bulk resistivity of the GO flake. However, if we assume that the bulk resistance is ohmic, we can obtain an upper limit. The dynamic resistance of the device is the sum of the dynamic resistances of two SBs and the bulk: $R=R_{SB1}+R_{SB2}+R_{bulk}$. We measured the $I$-$V$ of a Au/GO/Au structure up to the breakdown voltage of 17.5 V where it burned out (see \rfig{Burn}). Just prior to failure, the dynamic resistance was only 74 k$\Omega$, which sets the upper limit of the bulk resistivity at about 74 k$\Omega$/sq (aspect ratio $\sim$ 1). The doping density calculated from the slope of the $\ln(I)-V^{1/4}$ plot is about $9.9\times10^{10}$ cm$^{-2}$, and the mobility of this flake is therefore at least 850 cm$^2$/Vs.

We also succeeded in oxidizing both patterned and unpatterned EG chips \cite{supple}. Unpatterned EG chips were oxidized by Hummers method \cite{Hummers1958}. The surface morphology ({\it cf.} \rfig{IV}b) as measured by AFM exhibited no apparent changes from before to after oxidation. GO formation was verified by its characteristic Raman signature \cite{Stankovich2007} and a resistivity increase by orders of magnitude $>10^4$. Likewise, several ribbons were patterned on an as-grown EG chip. Hydrogen silsesquioxane was spun on the sample and e-beam patterned to produce rectangular windows over the central portions of the ribbons. The sample was subsequently oxidized. The resultant EG/GO metal-semiconductor-metal device is completely off, even at bias voltages up to 60 V, suggesting a large SB height (as for the GO bilayer flake, \rfig{IV}a). However, after subjection to e-beam irradiation (30 keV), the devices displayed a nonlinear $I$-$V$, which again is well described by \req{current} and (\ref{barrierlowering}). Because the devices are made from a continuous sheet of EG, impurities and interface states are essentially excluded. Hence, other than the SB, an insulating tunnel-barrier layer is unlikely to exist. For all three types of junctions, i.e. Au/GO, EG/GO and EG/GO (oxidized {\it in situ}), the $I$-$V$ is described by the same equations, strongly supporting our conclusion that the Schottky effect dominates the transport through those junctions. Further note that the Schottky barriers were found to be $\leq$ 0.5 eV after e-beam exposure. These significantly reduced SB heights indicate that e-beam treatment can be used to locally adjust the band gap, consistent with the known deoxidation of graphene oxide by electron beam exposure \cite{Galuska1988}.

In summary, we successfully produced all-graphene metal-semiconductor-metal devices. The $I$-$V$ characteristics of the device are explained by thermionic emission over a Schottky barrier. The barrier height is found to be as large as 0.7 eV, which indicates a band gap of at least this value in GO. The mobility of GO is larger than 850 cm$^2$/Vs, hence in the range suitable for room temperature electronics. Further tuning the band gap has been achieved by changing the degree of oxidation both by thermal curing and by e-beam irradiation.

The authors would like to thank T.~E. Mallouk and N.~I. Kovtyukhova for providing GO suspensions. This research was supported by NSF (NIRT grant No.0404084 and MRI grant No. 0521041), NERC, Intel, and CNRS. F. Ming acknowledges a Ph.D fellowship from IBM.

\end{document}